\documentclass[12pt]{article}

\usepackage{epsfig}

\begin{document}

\begin{center}
{\Large\bf The most powerful particles
\vspace*{3mm} \\ in the Universe: a cosmic smash}
\end{center}

\vspace*{1mm}

\begin{center}
Wolfgang Bietenholz \\
Instituto de Ciencias Nucleares \\
Universidad Nacional Aut\'onoma de M\'exico (UNAM) \\
A.P.\ 70-543, C.P.\ 04510 Distrito Federal, Mexico
\end{center}

\vspace*{3mm}

\noindent
{\bf This year we are celebrating 101 years since the discovery of
{\em cosmic rays.} They are whizzing all around the Universe, and they 
occur at very different energies, including the highest particle 
energies that exist.
However, theory predicts an abrupt suppression (a ``cutoff'')
above a specific huge energy. This is difficult to verify,
the measurements are controversial, but it provides a unique opportunity 
to probe established concepts of physics --- like Lorentz Invariance --- 
under extreme conditions. If the observations will ultimately
{\em contradict} this ``cutoff'', this could
require a fundamental pillar of physics to be revised.}

\section{Discovery of cosmic rays and air showers}

Throughout our lives we are surrounded --- and penetrated ---
by various types of radiation. Mankind was already aware of that 
in the beginning of the 20$^{\rm th}$ century, when 
instruments (like the electroscope) were
developed to detect ionizing radiation, and sources
inside the Earth (like the alkaline metal radium) were identified.
But does all the radiation around us originate from the Earth ?

If this was the case, the radiation intensity should decrease
rapidly with the height above ground.
In 1910 the German Jesuit
Theodor Wulf performed tests on top of the Eiffel tower, 
but they did {\em not} confirm the expected decrease. People
criticized, however, that the presence of tons of metal might 
have affected his results. More stringent experiments
were done on balloons; in particular in 1912 the
Austrian scientist Viktor Hess observed in seven 
balloon journeys that the ionizing radiation decreases only 
mildly up to a height of about 2000 m above ground, but as he 
rose even higher (up to 5350~m) it gradually {\em increased} 
again. He interpreted this observation correctly: 
significant radiation must come from outside the Earth. 
Comparing data taken at day and night, and during an eclipse, he also 
concluded that the sun cannot be a relevant source of these
{\em ``cosmic rays''}.
\begin{figure}
\begin{center}
\includegraphics[angle=0,width=.4\linewidth]{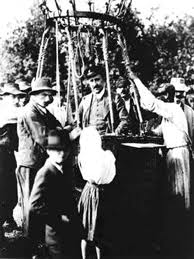}
\hspace*{8mm}
\includegraphics[angle=0,width=.4\linewidth]{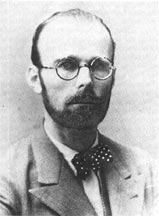}
\end{center}
\vspace*{-5mm}
\caption{\it{Viktor Hess (on the left) and Pierre Auger 
(on the right), the men who discovered the cosmic rays and
the air showers, respectively.}}
\vspace*{-3mm}
\end{figure} 

\begin{figure}
\begin{center}
\includegraphics[angle=0,width=.85\linewidth]{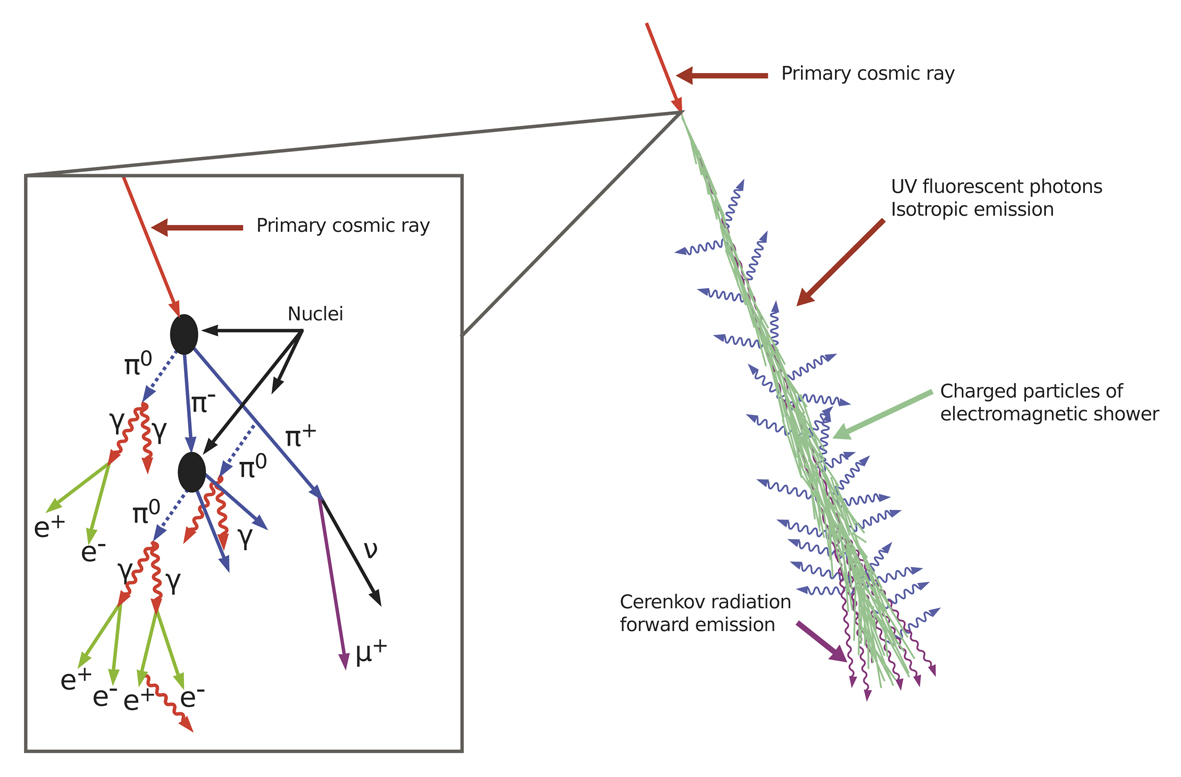}
\end{center}
\vspace*{-5mm}
\caption{\it{Illustration of an air shower. 
We recognize the so-called fluorescence light (UV or bluish),
and the generation of light particles named pions ($\pi$), 
which rapidly decay into even lighter leptons ($e, \, \mu, \, \nu$) 
and photons ($\gamma$).}} 
\label{airshower}
\vspace*{-3mm}
\end{figure}

As a further milestone, in 1938 the French physicist Pierre
Auger noticed that Geiger counters which were well separated
(by tens or hundreds of meters) often detected radiation
practically at the same time. He explained this effect as 
follows: a powerful cosmic ray particle (a ``primary particle'')
arrives from outer space and hits the terrestrial atmosphere.
Its violent collision with molecules of the air triggers
a cascade of ``secondary particles'', which we call an
{\em air shower.} Auger noticed that he had detected secondary 
particles belonging to the same air shower, which arrive on 
ground almost simultaneously. The formation of an air shower is 
illustrated in Figure \ref{airshower}.

It can be compared to a white ``primary'' billiard ball,
which hits (in the beginning of a game) a number of colored
balls, so its momentum is transferred and distributed
over numerous ``secondary'' balls. However, in an air shower
new ``balls'' are created in the collision, and in the subsequent
evolution; the more powerful the primary particle,
the more secondary particles emerge.

By analyzing his data taken at sea level and in the Swiss alps,
Auger conjectured that some primary particle energies should 
be at least of the order of $10^{15}~{\rm eV}$.\footnote{An
electron volt (eV) is the energy that it takes to displace
an object with the electric charge of an electron against
the voltage of 1 V. It is a very small unit of energy, which
is commonly used in quantum physics. We can convert it to
macroscopic units as follows: 
$6.2 \cdot 10^{18}~{\rm eV} = 1 ~ {\rm J} = 
1 ~ {\rm kg~m^{2} / s^{2} }$, and 
$10^{18}$ means $1\, 000\, 000\, 000\, 000\, 000\, 000$ (18 zeros).}

The kinetic energy is a measure for how much work is
needed to accelerate an object from rest to a given speed.
For comparison, a table tennis ball with a speed of 34~cm/s 
has the same kinetic energy, $10^{15}~{\rm eV}$, but a
$5.4 \cdot 10^{23}$ times larger mass, if we assume the
primary cosmic ray particle to be a proton 
(we recall that the tiny nucleus of a hydrogen atom consists 
of a proton).

\section{The profile of the cosmic flux}

Today we know about cosmic rays in the energy range of 
$E \approx 10^{9} \dots 10^{20} ~ {\rm eV}$.
Up to now we only know of their existence, but hardly 
anything about their origin.\footnote{Radiation 
at lower energy is also present, and here the sun does contribute 
significantly, but we do not denote that as ``cosmic rays''.}
The top energy, about $10^{20} ~ {\rm eV}$, is 100\,000 times larger 
than Auger's estimate. This corresponds to a
tennis ball with 85 km/h, or table tennis ball with 392 km/h
(for comparison, the hardest smashes in professional table tennis 
games attain about 100 km/h).

\begin{figure}
\begin{center}
\includegraphics[angle=0,width=.9\linewidth]{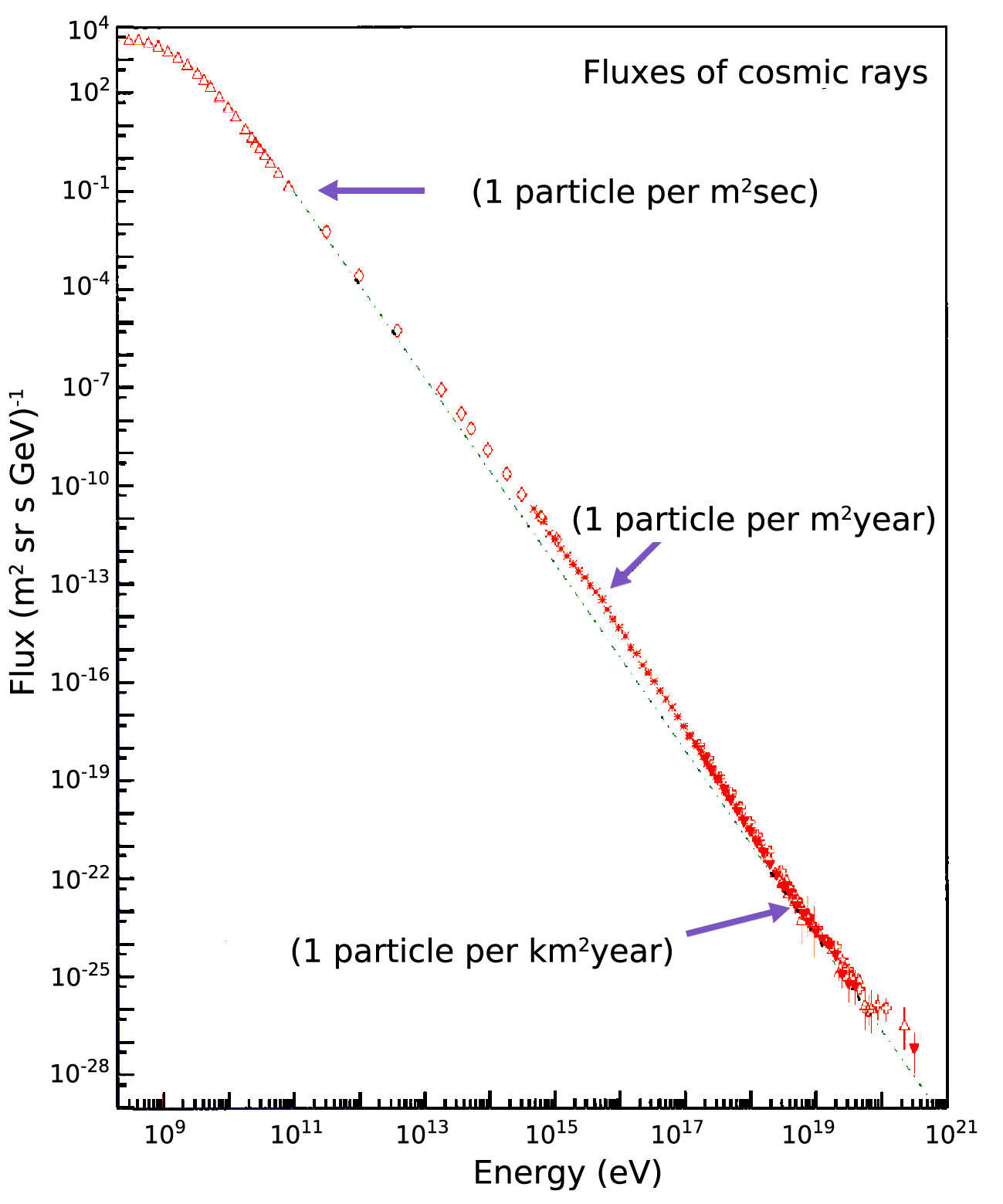}
\end{center}
\caption{\it The flux of cosmic rays as a function of the energy. 
Over a very broad energy interval it falls off approximately proportional 
to $1/ E^{3}$ (dashed line). 
Around $E = 6 \cdot 10^{19}~{\rm eV}$ an abrupt flux reduction is 
predicted; this is the GZK cutoff.}
\vspace*{-6mm}
\label{flux}
\end{figure}

One assumes the high energy cosmic rays to consist to about $90 \ \%$ 
of protons, and to $9 \ \%$ of helium nuclei.
They are whizzing all around the Universe, in all directions,
at any time. We may wonder how many there are, {\it i.e.}\
how many cosmic ray particles cross a given area per time.
This is what we denote as the cosmic {\em flux.}
Over the entire energy range, this flux follows closely a curve
proportional to $1 / E^{3}$, see Figure \ref{flux}.
So if we double the energy at which we measure the flux,
it will decrease by a factor of 8.
The validity of such a simple rule over such a huge range is
very remarkable; the ratio between its lowest and highest
energies corresponds to the ratio between the size of a human
body and our distance from the sun. This is
impressive, but the reason for this rule is not understood.

Around $10^{12} ~{\rm eV}$ the flux is $10$ primary
particle per minute and m$^{2}$ (convenient for measurements), 
but as we approach the upper end of the known spectrum, 
say between $10^{18}$ and $10^{19}~{\rm eV}$, we are
left with only $1$ primary particle per year and km$^{2}$;
here the detection takes a large area, and a lot of patience.
But what happens at {\em even higher energy,} does the flux continue
with the same $1/E^{3}$ power law and we just haven't measured
it well so far ?

\section{From the Cosmic Microwave Background to the
prediction of the ``GZK cutoff''}

In 1965 A.\ Penzias and R.\ Wilson discovered (accidentally)
the Cosmic Microwave Background (CMB), which is a relic of the Early
Universe: its photons (the quanta of electromagnetic radiation)
decoupled some $380 \, 000$ years after the Big Bang, when the
Universe only had $0.0028 \%$ of its age today.
This photon radiation cooled down ever since, 
so at present the CMB --- and therefore the 
Universe --- has a temperature of 2.73 K.\footnote{The
absolute temperature minimum is $0 ~ {\rm K} = -273.15~^{0}
{\rm C}$, hence the CMB temperature corresponds to 
$-270.42~^{0}{\rm C}$.}
This means that one cm$^{3}$ contains in average 411 CMB photons, 
with a mean wave length of 1.9~mm, which corresponds to a tiny
energy of $0.0006~{\rm eV}$.\\

One year later, this discovery led to an epoch-making theoretical
work, independently by K.\ Greisen at Cornell University 
(state of New York), and by G.T.\ Zatsepin and V.A.\ Kuz'min 
at the Lebedev Institute (Moscow) \cite{GZK}:
they (we denote them as GZK) predicted the cosmic ray spectrum 
to have a ``cutoff'' around 
$E_{\rm GZK} = 6 \cdot 10^{19}~{\rm eV}$, {\it i.e.}\ they
predicted that the flux above $E_{\rm GZK}$ should nearly vanish.
These two papers have a renowned status, although they were 
both short, with hardly any formulae, but with a superb idea.

\begin{figure}
\begin{center}
\includegraphics[angle=0,width=.29\linewidth]{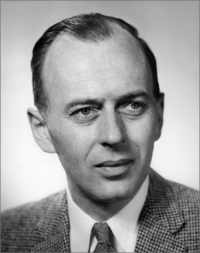}
\includegraphics[angle=0,width=.32\linewidth]{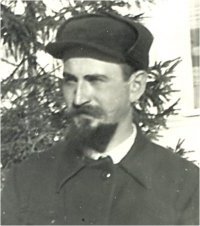} 
\includegraphics[angle=0,width=.32\linewidth]{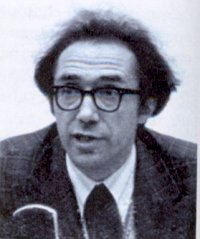}
\end{center}
\caption{\it {\em From left to right:} Kenneth Greisen (1918 -- 2007),
Georgiy Zatsepin (1917 -- 2010) and Vadim Kuz'min (born 1937),  
the theoretical physicists who predicted in 1966 the ``GZK cutoff'' 
for the cosmic ray spectrum.}
\end{figure} 

\begin{figure}
\begin{center}
\includegraphics[angle=0,width=1.\linewidth]{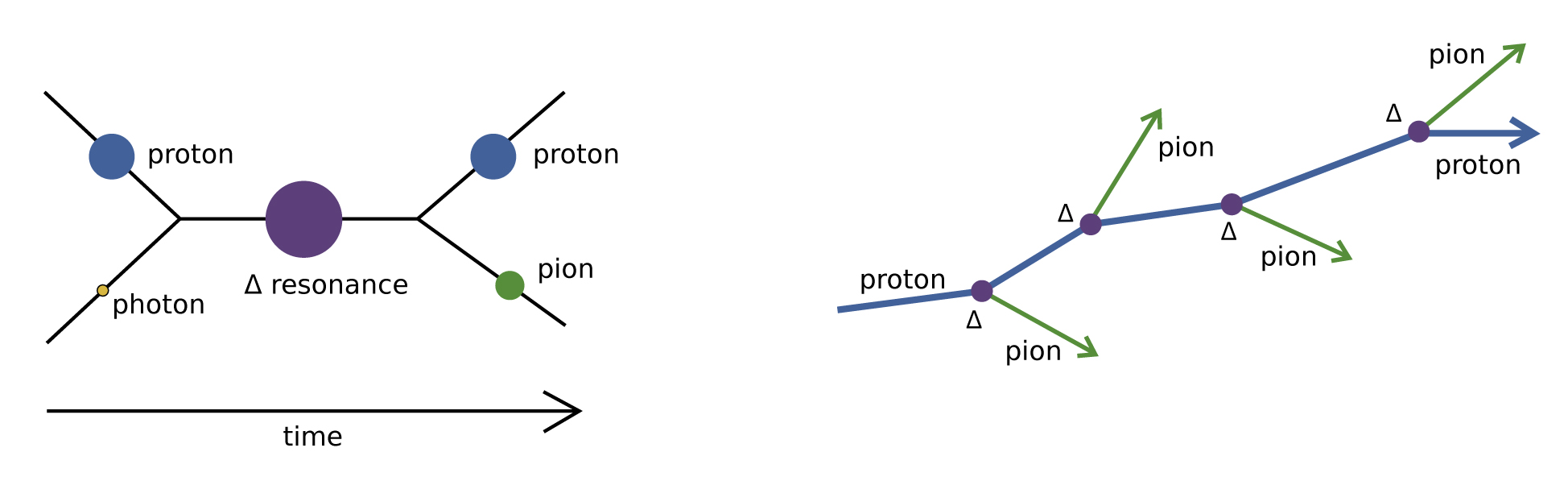} 
\end{center}
\vspace*{-3mm}
\caption{\it {\em On the left:} The scheme of photopion production 
due to the collision of an ultra high energy proton with a 
CMB photon. {\em On the right:} Trajectory of a super-GZK
proton (a proton with energy above $E_{\rm GZK}$)
through the CMB, suffering energy attenuation due to
repetitive photopion production.}
\label{GZK}
\end{figure}

Their point was that the scattering
of protons with photons can generate a heavier particle, which we now
denote as a ``$\Delta$ resonance'' (similar to a violin string
vibrating above its ground frequency).
It is short-lived, and its decay reproduces
the proton, along with an (aforementioned) lighter particle
called pion (``photopion production''), as illustrated in
Figure \ref{GZK} on the left.\footnote{Photopion production can 
also occur in a less direct way, where $\Delta$ first decays into
a pion and a neutron, and the latter is converted subsequently 
into a proton through $\beta$-decay. These two channels together
cover 99.4 \% of the $\Delta$ decays.}
The pion carries away 
part of the energy, typically about $20 ~ \%$.\footnote{For proton 
energies well above  $E_{\rm GZK}$ also higher resonances are possible, 
where the decay may yield several pions, so that the proton loses 
even more energy.\label{multipion}}
$E_{\rm GZK}$ is just the threshold energy for a cosmic proton to create 
such a $\Delta$ resonance when hitting head-on a (relatively energetic) 
CMB photon. So if a proton with an even higher energy travels
through the Universe, it will undergo this process again and again,
and lose energy each time, until it drops below $E_{\rm GZK}$.
This step-wise attenuation is sketched in Figure \ref{GZK} on the right.

As a rough picture, we could imagine a car driving very fast,
above the speed limit. As a consequence it touches obstacles
here and there, say without a bad accident, but losing
speed each time. This is repeated until the car has slowed down 
below the speed limit, then it does not suffer from further 
accidents anymore. So at the end of a long road, all cars will
necessarily arrive with an allowed speed.

Considering the photon density that we mentioned above, 
and the target area (``cross section'') for a proton-photon collision
leading to a $\Delta$ resonance (around $10^{-28} ~ {\rm cm}^{2}$),
the mean free path length --- between two such collisions ---
for a proton just above $E_{\rm GZK}$ is 
around $15 ~ {\rm Mpc}$.\footnote{One parsec (pc) is a standard 
length unit in astronomy, which corresponds to 
$3.1 \cdot 10^{16} ~ {\rm m}$, or 3.3 light-years.
$1 ~ {\rm Mpc} = 10^{6} ~ {\rm pc}$ 
means one million of parsecs.} If the initial proton energy is much 
higher, the energy attenuation is much more rapid, since photopion 
production is more frequent, and the proton loses more energy each 
time. In that case also the emission of several pions is 
possible (cf.\ footnote \ref{multipion}).

One concludes that protons can travel maximally about 
$L_{\rm max} = 100 ~ {\rm Mpc}$ with super-GZK energy,
$ E > E_{\rm GZK}$. If 
the primary ray consists of heavier nuclei, this maximal distance 
is shorter, because such a nucleus tends to break apart under 
scattering, such that its fragments lose even more energy.

$L_{\rm max}$ is a long distance compared to the radius of 
our galactic plane of about $0.015 ~ {\rm Mpc}$, but it is short 
compared to the radius of the visible Universe, which is 
around $14 \, 000 ~ {\rm Mpc}$.
So if sources of ultra high energy cosmic rays are spread 
homogeneously in the Universe, the flux that we observe on 
Earth should have a strong extra suppression as the energy 
exceeds $E_{\rm GZK}$, pushing its intensity well below
the extrapolated $1/E^{3}$ rule. This is not a strict cutoff 
--- although it is referred to as the ``GZK-cutoff'' --- but it 
is an interesting and explicit prediction.
Its verification is a tough challenge for our best observatories.

\section{Observations of ultra high energy cosmic rays in the
20$^{\rm th}$ century}

In 1963, already before this prediction was put forward, one 
super-GZK event with an estimated energy of 
$10^{20} ~{\rm eV}$ was reported by J.\ Linsley,
based on an air shower detected in the desert of
New Mexico (USA). This issue attracted interest world-wide.
K.\ Greisen expressed his surprise about that, and added 
that he did not expect any events at even higher energy.

Nevertheless, in 1971 another super-GZK event was observed in 
Tokyo, this time with even higher energy. This inspired the 
construction of a large observatory near the Japanese town
Akeno, which is called AGASA (Akeno Giant Air Shower Array).
Until the end of the last century AGASA dominated the
world data about ultra high energy cosmic rays.
It recorded about two dozens of new super-GZK events, and 
compatibility with the $1/E^{3}$ rule also beyond $E_{\rm GZK}$,
{\em in contrast to the prediction} \cite{AGASA}.
This picture was essentially supported by somewhat
smaller installations in Yakutsk (Russia) and 
Haverah Park (England). The world record is generally considered a
primary particle with $3 \cdot 10^{20} ~{\rm eV}$, reported in 
1991 by the Fly's Eye detector in Utah (USA), 
which was constructed like the compound eye of an insect.

We are lucky that such ultra high energy rays form air shower 
about 15 km above ground, so that their energy is spread over 
many secondary particles, rather than hitting us directly.
In macroscopic terms, this energy world record corresponds 
to $48 ~ {\rm J}$, and to the kinetic energy of a tennis ball 
with a considerable speed of 147 km/h. If the ball drops in vacuum
from a tower of 85~m height, it will hit the ground with this
speed (with air resistance it never gets that fast).  
This is still not the maximal speed in a professional tennis 
game; the second service of Novak Djokovic --- currently the tennis 
star number one --- is around 160 km/h, and his first service sometimes
exceeds 200 km/h. According to AGASA even that energy should 
be reached by single cosmic protons.

\section{Doubts about a fundamental law of physics~?}

Is this true, despite the stringent theoretical
argument by Greisen, Zatsepin and Kuz'min~?
This scenario fascinates physicists, since it would be
a clear indication of a phenomenon at tremendous energy,
which is incompatible with our established theories,
so its explanation would require {\em new physics}.
Numerous ideas were elaborated to explain the
possible failure of this prediction. The most prominent
approach is a {\em violation of Lorentz Invariance,}
see {\it e.g.}\ Ref.\ \cite{ColGla}, and Ref.\ \cite{WB}
for a recent review.

\begin{figure}
\begin{center}
\includegraphics[angle=0,width=.55\linewidth]{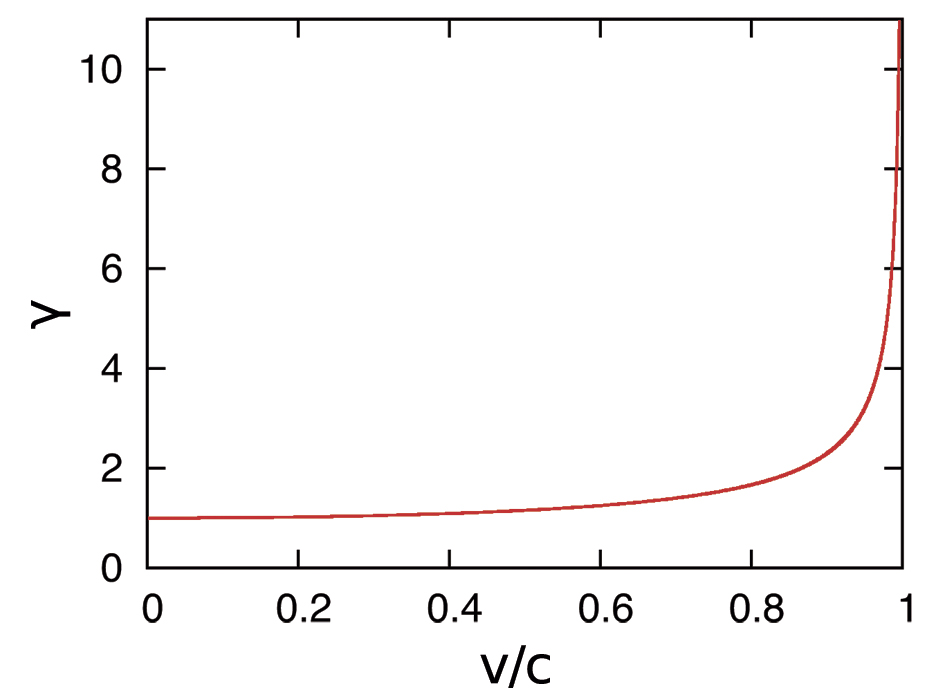}
\end{center}
\vspace*{-5mm}
\caption{\it The boost factor $\gamma$ as a function
of the relative speed $v$ in a Lorentz transformation.
This speed is displayed in units of $c$, the speed of light.
$\gamma = 1 / \sqrt{1 - v^{2}/c^{2}}$ is close to $1$ if
$v \ll c$, but it diverges as $v$ approaches $c$.}
\label{gamma}
\end{figure}

Lorentz Invariance means that observers moving with constant 
speed relative to each other --- for instance living on
different space stations --- perceive the same laws of 
physics, hence there is no ``preferred'' reference frame.
This is one of the most fundamental pillars of our
physical concepts.
The observed quantities are transformed according
to simple formulae of Einstein's Theory of Relativity 
(``Lorentz transformation''). In particular the speed of 
light must be invariant.\footnote{This property is different from the 
non-relativistic ``Galilei transformation'', which was used
until the beginning of the 20$^{\rm th}$ century.}
A Lorentz transformation is 
characterized by a boost factor called $\gamma$, which
translates for instance a length, a time period or an energy
as it is perceived by the two observers. $\gamma$
grows monotonously with the relative speed between the observers,
{\it i.e.}\ a faster speed implies a larger $\gamma$.
It goes towards infinity when this relative velocity approaches 
the speed of light, see Figure \ref{gamma}. So in this case
the perceptions of the two observers --- {\it e.g.}\ of the
length of a given object --- are drastically different.

The validity of Lorentz Invariance is very well tested and
confirmed with our most powerful particle accelerators
up to $\gamma$-factors around $10^{5}$, in particular due to the 
Large Electron-Positron Collider (LEP), which was used from
1989 to 2000 at CERN in Geneva.\footnote{The now operating
Large Hadron Collider (LHC, also at CERN) attains even higher 
energies, but for particles and nuclei which are much
heavier than the electron and positron, so LEP still 
holds the {\em speed} world record in laboratories.\label{LHC}}
This excludes ``preferred reference frames'' at that level. 
Here the observers move relative to each other
with 99.999999995 \% of the speed of light.
(No massive object can ever attain exactly the speed of 
light --- that would require an infinite amount of energy.)

What does this imply for ultra high energy cosmic rays ?
So far we have tacitly assumed Lorentz Invariance to hold.
For example, the mean free path length of a cosmic super-GZK proton
--- before performing the next photopion production --- that
we mentioned in Section 3 (some 15 Mpc) is based on measurements of 
the proton-photon cross section in laboratories. Actually
our accelerators cannot provide such tremendous proton energies.
Even the most powerful accelerator in history,
the LHC (referred to in footnote \ref{LHC})
stays far below that. One may use, however, protons 
at rest and expose them to a photon beam of about 
$200 ~ {\rm MeV} = 2 \cdot 10^{8} ~ {\rm eV} $, which is 
equivalent, {\em if} we assume Lorentz Invariance to hold.
This is rather easy for experimentalists, it was done
already in the 1950s, so Greisen, Zatsepin and Kuz'min 
could refer to the result. Also the computation of the
energy loss of an ultra high energy proton under photopion
production that we mentioned in Section 3 (about 20 \%) is based on 
Lorentz Invariance. 

However, the transformation factor between a proton
at rest, and with an energy around $E_{\rm GZK}$, amounts to
$\gamma = E_{\rm GZK}/ m_{\rm p} \approx 10^{11}$, 
{\em far} beyond the boost factors that have ever been tested 
($m_{\rm p}\simeq 9.38 \cdot 10^{8} ~{\rm eV}$ is the proton mass). 
A proton with energy $E_{\rm GZK}$ moves with 
99.999999999999999999995 \% of the speed of light.
So could Lorentz Invariance be an excellent approximation up 
to $\gamma \approx 10^{5}$, which still requires
some modification at much larger $\gamma$ values~?

The possible {\em absence} of the GZK cutoff for cosmic rays
could be a hint for this scenario.
This question has to be addressed experimentally,
and it is a fascinating opportunity to probe
our established theory under truly extreme conditions,
which are by no means accessible in our laboratories.

\section{The phenomenological situation today}

In the beginning of the 21$^{\rm st}$ century the phenomenological
situation changed, when the HiRes (High Resolution) Observatory 
in Utah (USA) started to dominate the world data \cite{HiRes}. 
Its results favor the conclusion {\em opposite} to AGASA, {\it i.e.}\ 
the ``boring scenario'' where the GZK cutoff is confirmed, 
no new physics 
is needed, and Djokovic's service is not
challenged by cosmic protons. In the contrary, this would
provide indirect evidence {\em for} the validity of
Lorentz Invariance, even at such tremendous speed transformations.

How could this discrepancy with AGASA and other observatories
occur~? A possible explanation is that they used different
techniques: AGASA, Ya\-kutsk and Haverah Park detected on ground
secondary particles of powerful air showers. As a rough 
rule, such a shower involves (in its
maximum) about 1 particle for each GeV of the
primary particle ($1 ~{\rm GeV} = 10^{9}~{\rm eV}$), 
so that a $10^{20} ~{\rm eV}$ proton can
give rise to a shower of up to $10^{11}$ secondary particles
(in this respect, the colored billiard balls cannot compete).
After detecting some of these secondary particles, fast computers
and sophisticated
numerical methods are used to reconstruct the most likely
point where the shower took its maximum.
That indicates the nature of the primary particle (or nucleus) ---
the heavier it is, the higher the shower maximum.
This numerical reconstruction of the air shower evolution
also provides an estimate for 
the primary particle energy --- obviously with some uncertainty.

On the other hand, HiRes monitored a weak bluish or ultraviolet
``fluorescence light'' (see Figure \ref{airshower}).
It originates in nitrogen molecules in the air, 
which are excited by an air shower, and which emit this light when
returning to their ground state. The virtue 
is that the shower is observed at an early stage, so it does not
need to be reconstructed afterwards numerically. On the other hand, 
this observation is only possible in nights without clouds and 
without much moon shine, hence it provides only modest statistics.\\

In order to settle this controversy, the {\em Pierre Auger Observatory} 
in Argentina now combines {\em both} techniques \cite{PAO}. On ground
1600 water tanks detect secondary particles and capture many
high energy cosmic rays. Their installation was completed
in 2008, and its data set now clearly exceeds the 
previous world statistics. Moreover 24 fluorescence telescopes
search for ``golden events'' which are observed by both systems;
they are very helpful in verifying the estimate for the primary 
particle energy. Thus the systematic error is around $22 ~ \%$,
which is harmless in this business, where one deals with 
magnitudes.
\begin{figure}
\begin{center}
\includegraphics[angle=0,width=.45\linewidth]{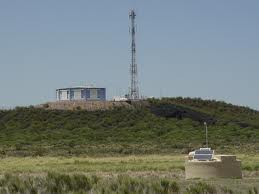}
\includegraphics[angle=0,width=.45\linewidth]{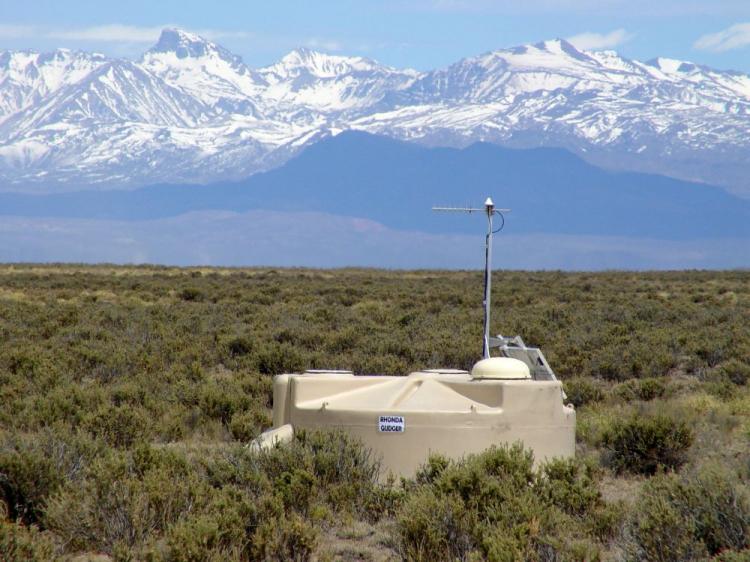}
\vspace*{3mm} \\
\includegraphics[angle=0,width=.9\linewidth]{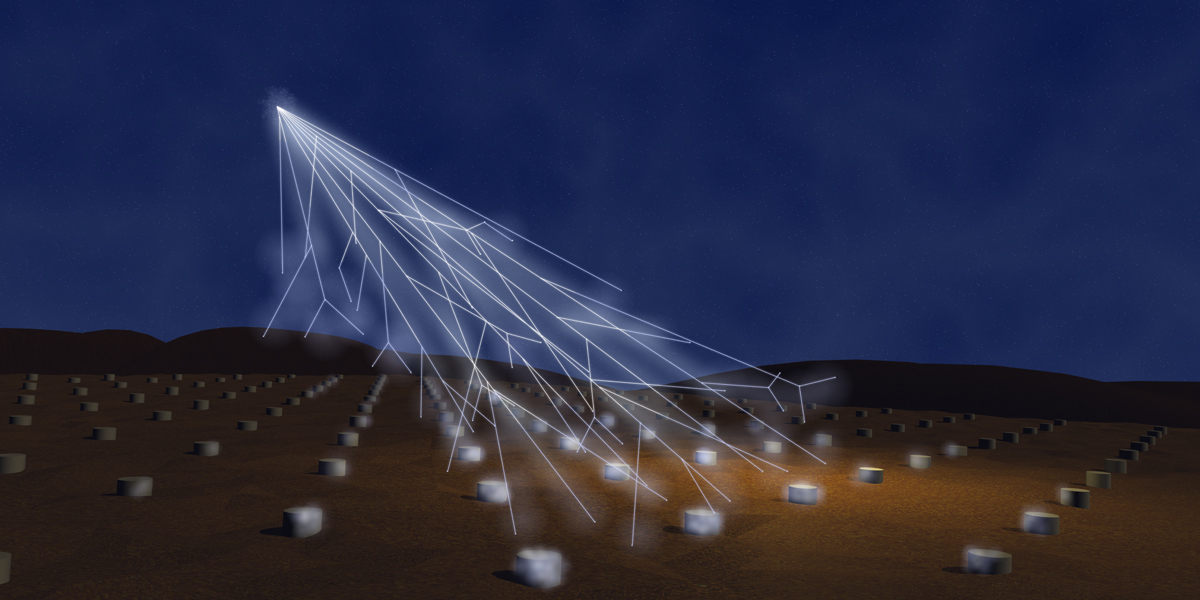}
\end{center}
\vspace*{-2mm}
\caption{\it Images of the Pierre Auger Observatory in the ``Pampa 
Amarilla'' (Yellow Prairie) of the province of Mendoza, Argentina. 
It is now the dominant observatory in the
search for ultra high energy cosmic rays. 
The collaboration involves almost 500 scientist in 19 countries.
\newline
{\em Above, on the left:} one of the 24 fluorescence telescopes
(on the hill), with cameras monitoring the weak bluish light that 
an air shower emits. 
{\em Above, on the right:} a cylindric tank, dark inside, with 
12\,000~l of water, able to detect secondary particles. 
1600 such tanks are spread over an area of 3000 km$^{2}$, on a
triangular grid with spacing 1.5~km, in order 
to capture multiple secondary particles of a powerful air shower.
{\em Below:} an artistic illustration.}
\label{PAOpics}
\end{figure} 
Up to now the Pierre Auger Observatory has accumulated a lot of 
data, in particular it has identified over 100
primary particles with energies close to or above $E_{\rm GZK}$. 
However, even with these new data the statistics is still 
not sufficient for a conclusive answer to the question
if there really is a GZK cutoff for the energy of cosmic rays.
We add a short discussion in the Appendix.

\section{Outlook}

New experiments are in preparation, such as
JEM-EUSO (Japanese Experiment Module -- Extreme Universe Space 
Observatory) or OWL (Orbiting Wide-angle Light-collectors):
now the idea is to observe the air shower formation from above, 
{\it i.e.}\ from satellites, which should provide more precise
information. They will monitor the showers from the
beginning, without being obstructed by clouds.

Hopefully this will at last answer the outstanding question 
about the existence of the GZK cutoff, which has fascinated 
scientists for almost half a century \cite{WB}. Then we should
finally know whether or not our established physical framework --- 
with Lorentz Invariance as a cornerstone --- needs to be revised,
and whether or not cosmic protons can 
compete with services, or even smashes, in a professional
tennis game. \\

\noindent
{\small\it I am indebted to Alberto G\"{u}ijosa, Marco Panero and 
Uwe-Jens Wiese for useful comments, 
and to Aline Guevara for her help with the figures.
A shorter version of this article in Spanish, with co-author
Ang\'{e}lica Bahena Blas, is accepted for publication in
{\em CIENCIA, Revista de la Academia Mexicana de Ciencia.} 
It is available online: \
{\em http://www.revistaciencia.amc.edu.mx/online/15$_{-}$779$_{-}$Particulas.pdf}}

\appendix

\section{The flux of ultra high energy cosmic rays}

The Pierre Auger Observatory operated in part since 2003.
In 2007 it released preliminary results, which supported
the scenario advocated by HiRes: the GZK cutoff seemed to be 
confirmed, although a number of new super GZK events were found.

\begin{figure}
\begin{center}
\includegraphics[angle=0,width=.7\linewidth]{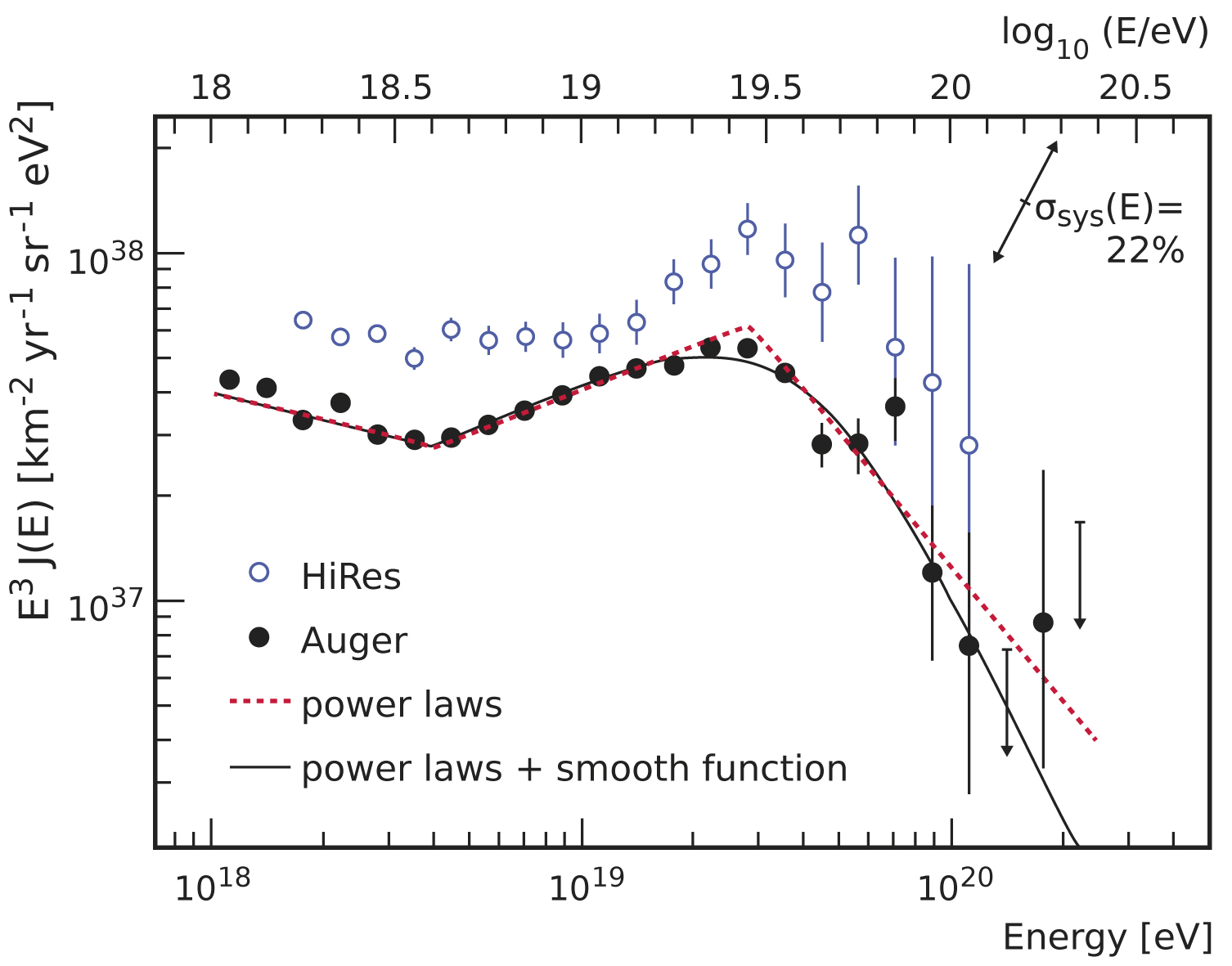}
\end{center}
\vspace*{-5mm}
\caption{\it The flux of cosmic rays in the ultra high energy 
regime, multiplied by the factor $E^{3}$ (where $E$ is the energy), 
according to the Observatories HiRes and Pierre Auger \cite{PAO}.}
\label{PAO2010}
\end{figure}

However, meanwhile the Pierre Auger Observatory has accumulated
more and more statistics, and the conclusion about the ultra high
energy cosmic flux is not obvious. Some excess 
--- compared to the $1/E^{3}$ rule ---
is clearly observed just above $4 \cdot 10^{18}~{\rm eV}$,
see Figure \ref{PAO2010}. Above $3 \cdot 10^{19}~{\rm eV} \approx E_{\rm GZK}$ 
the flux drops quite sharply, 
which {\em can} be regarded as further evidence for the ``boring
scenario''. However, if we extrapolate the flux from $E <
4 \cdot 10^{18}~{\rm eV}$ into the super-GZK regime, it is well
compatible with the data. Hence this excess could also be viewed
as a limited pile-up (as it also occurs at lower energies,
see Figure \ref{flux}), while the extended power law
might still be in business.

Therefore, even with the new data by the Pierre Auger Observatory, the
statistics is still too poor for a conclusive answer to the question
if there really is a GZK cutoff for the energy of cosmic rays.

\end{document}